# Realizing the insulator-to-metal transition in Se-hyperdoped Si via non-equilibrium material processing

Fang Liu,[1,2] S. Prucnal,[1] Y. Berencén,[1] Zhitao Zhang,[1] Ye Yuan,[1,2] Yu Liu,[1] R. Heller,[1] R. Böttger,[1] L. Rebohle,[1] W. Skorupa,[1] M. Helm,[1,2] and Shengqiang Zhou[1]

[1]Helmholtz-Zentrum Dresden-Rossendorf, Institute of Ion Beam Physics and Materials Research, Bautzner Landstr. 400, 01328 Dresden, Germany

[2]Technische Universität Dresden, 01062 Dresden, Germany

E-mail: f.liu@hzdr.de

**Abstract.** We report on the insulator-to-metal transition in Se-hyperdoped Si layers driven by manipulating the Se concentration via non-equilibrium material processing, i.e. ion implantation followed by millisecond-flash lamp annealing. Electrical transport measurements reveal an increase of carrier concentration and conductivity with increasing Se concentration. For the semi-insulating sample with Se concentrations below the Mott limit, quantitative analysis of the temperature dependence of conductivity indicates a variable-range hopping mechanism with an exponent of $s = 1/2$ rather than $1/4$, which implies a Coulomb gap at the Fermi level. The observed insulator-to-metal transition is attributed to the formation of an intermediate band in the Se-hyperdoped Si layers.

# 1. Introduction

Recently, many efforts have been focused on exploring intermediate band (IB) materials [1]. The IB permits photons with a low energy below the band gap of the semiconductor to be absorbed, yielding a high optical absorbance in the low energy part of the solar spectrum [2-5]. For deep impurities, like S and Se in Si with a binding energy of around 300 meV [6, 7], the insulator-to-metal transition is a direct evidence for the formation of an IB in host semiconductors [8, 9]. The insulator-to-metal transition controlled by doping occurs at a critical impurity concentration [10]. The growing electron density gives rise to the screening of the Coulomb potential and results in metallic-like conductivity when the doping concentration is above the Mott limit [8, 10]. Most of the insulator-to-metal transitions are driven by doping semiconductors with shallow dopants, due to their low critical doping concentrations of around $10^{18}$ cm$^{-3}$ [11]. In contrast, deep-level dopants provide more tightly bound electrons, but need a higher critical doping concentration of $10^{19} \sim 10^{20}$ cm$^{-3}$ [12]. However, under thermal equilibrium conditions the solid solubility limits of deep-level dopants are only about $10^{16}$ cm$^{-3}$ in silicon, which are much lower than those for shallow dopants, viz. around $10^{21}$ cm$^{-3}$ [13]. In order to overcome the solid solubility limits of deep level dopants in Si, non-equilibrium techniques are needed. Ion implantation is a non-equilibrium technique which can introduce dopants into semiconductors beyond the solid solubility limit. To electrically activate the dopants and to repair the lattice damage induced by energetic ions, a post-implantation annealing is necessary. Both pulsed laser annealing in the nanosecond range [8, 14] and furnace annealing in the minute one [15, 16] have been used to activate S and Se dopants in Si introduced by ion implantation. However, the long-time process required for furnace annealing or even few seconds for rapid thermal annealing will cause large possibilities for impurity diffusion, resulting in a low substitutional fraction [15] or a deactivation of dopants [17]. During pulsed laser annealing the surface temperature exceeds the melting point, subsequently leading to liquid-phase epitaxial regrowth. The main problem of liquid-phase epitaxy is the redistribution of impurities [18] and the cellular breakdown [19]. Alternatively, it has recently been demonstrated that millisecond range flash-lamp annealing (FLA) induces solid phase epitaxy which accounts for a complete recrystallization of the implanted layer at annealing temperatures below the melting point of substrate [20-22]. During this process, the impurities have more probability to be incorporated into the semiconductor lattice due to i) the high velocity of the solidification during recrystallization after FLA and ii) the low diffusion velocity of impurities in the solid phase [23, 24]. It has been shown that FLA is superior to laser annealing in preventing the surface segregation of dopants and in suppressing the cellular breakdown for semiconductors with high impurity concentration [25, 26]. The structural and electrical properties of Se-hyperdoped Si depending on different annealing parameters have been reported in Ref. 25.

In this work, Se-hyperdoped Si samples with different Se concentrations have been fabricated by ion implantation followed by FLA. The insulator-to-metal transition is shown to occur at a Se peak concentration of around $6.3 \times 10^{20}$ cm$^{-3}$. Variable-range hopping with an exponent of s = 1/2 is found to be the main electrical transport mechanism for the semi-insulating sample with Se concentrations below the Mott limit. The existence of a Coulomb gap at the Fermi level is also inferred.

## 2. Experimental

Semi-insulating (100) Si wafers with a thickness of 525 μm and a resistivity as large as $10^4$ Ω·cm were used for this investigation. The 4-inch wafers were implanted at energy of 60 keV with Se fluences (Φ) of $1.0 \times 10^{15}$, $2.5 \times 10^{15}$, and $5 \times 10^{15}$ cm$^{-2}$, which results in three different Se peak concentrations ($c_{pk}$) of $2.5 \times 10^{20}$, $6.3 \times 10^{20}$, and $1.3 \times 10^{21}$ cm$^{-3}$, respectively. $c_{pk}$ is calculated by the following equation: $c_{pk} = \frac{\Phi}{\sqrt{2\pi}\Delta R_P}$, where $\Delta R_P$ is the longitudinal straggle [14, 27]. The Se peak concentration is chosen to cover the range of the insulator-to-metal transition predicted by density function theory (DFT) [14], Quantum Monte Carlo (QMC) calculations [14] and hybrid functional method [28]. The as-implanted samples were annealed by FLA with fixed pulse duration of 1.3 ms and an energy density of around 20 J/cm$^2$ in N$_2$ ambient. Prior to the flash, the samples were preheated up to 400 °C for 30 s. The estimated peak temperatures at the sample surface were in the range of 1100-1200 °C. Hereafter, the samples will be denoted by their Se peak concentrations unless otherwise stated.

The Raman scattering spectra were obtained using a 532 nm Nd:YAG laser and a charge coupled device camera cooled with liquid nitrogen. The lattice location of Se in the Si matrix was analyzed by means of channeling Rutherford backscattering spectrometry (cRBS). cRBS measurements were performed with a collimated 1.7 MeV He$^+$ beam at 170° backscattering angle. The RBS spectra were measured in both random and channeling geometries. The electrical properties of Se-implanted samples were examined by the van der Pauw configuration using a commercial Lakeshore Hall System in the temperature range of 2.5–300 K.

## 3. Results

The regrowth quality of the Se-implanted layers was determined by Raman spectroscopy. Fig. 1 shows the Raman spectra of the as-implanted and FLA treated samples with different Se peak concentrations. The spectrum of the virgin Si is also included for comparison. The Raman spectrum of virgin Si exhibits a narrow peak at around 520 cm$^{-1}$ which corresponds to the transverse optical (TO) phonon mode of single-crystalline Si. A weak band at around 460 cm$^{-1}$ can be observed in the as-implanted sample, which indicates the formation of an amorphous silicon layer during ion implantation [29, 30]. The peak at around 520 cm$^{-1}$ of the as-implanted sample is from the Si substrate beneath the implanted amorphous layer. However, after FLA the weak band associated with the amorphous layer vanishes, while the intensity of the 520 cm$^{-1}$ Raman peak of crystalline Si restores to that of the virgin Si. The peak observed at 303 cm$^{-1}$ corresponds to the second transverse acoustic (2TA) phonon mode of single-crystalline Si. Thus, Raman results prove that the crystal order can be well restored using FLA.

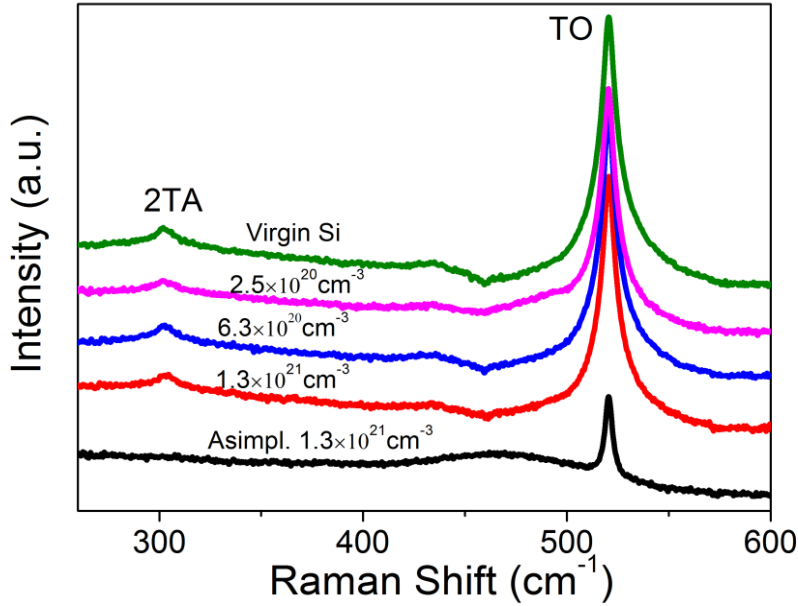

FIG. 1. Room-temperature Raman spectra of Se-implanted Si samples (Se concentration of $2.5 \times 10^{20}$, $6.3 \times 10^{20}$, and $1.3 \times 10^{21}$ cm$^{-3}$) annealed at 1.3 ms with an energy density of 20 J/cm$^2$. For comparison, the spectrum of the virgin Si is also shown.

In order to further analyze the crystalline properties of Se implanted Si samples after annealing, we performed cRBS measurements. Figure 2 shows the RBS spectra of the as-implanted, un-implanted and implanted samples followed by FLA. The channeling spectra were measured along the [100] axis for all samples. The minimum yield $\chi_{min}$ is defined as the ratio of the backscattering yield of the channeling and random spectra, which is an important indicator of the lattice quality [25]. If the implanted layer is fully amorphous, $\chi_{min}$ is 100%, while the $\chi_{min}$ value of 1–2% stands for a high-quality single crystal material. The cRBS spectrum of the as-implanted sample shows a broad damage band located between 900 and 970 keV. The $\chi_{min}$ for the as-implanted sample is calculated to be around 94.3%. The lattice was significantly damaged after Se implantation compared with a fully amorphous layer ($\chi_{min} \sim 100\%$). After FLA, however, the backscattering yield of all samples is similar to the virgin Si sample, which indicates that a high degree of lattice recrystallization has been obtained. To investigate the lattice location of the Se atom, we show the Se-related peak in Fig. 2. For the sample with the largest Se concentration, the channeling spectrum is significantly reduced in backscattering yield compared to the random spectrum, indicating that around 70% of Se atoms are on a substitutional lattice location [25].

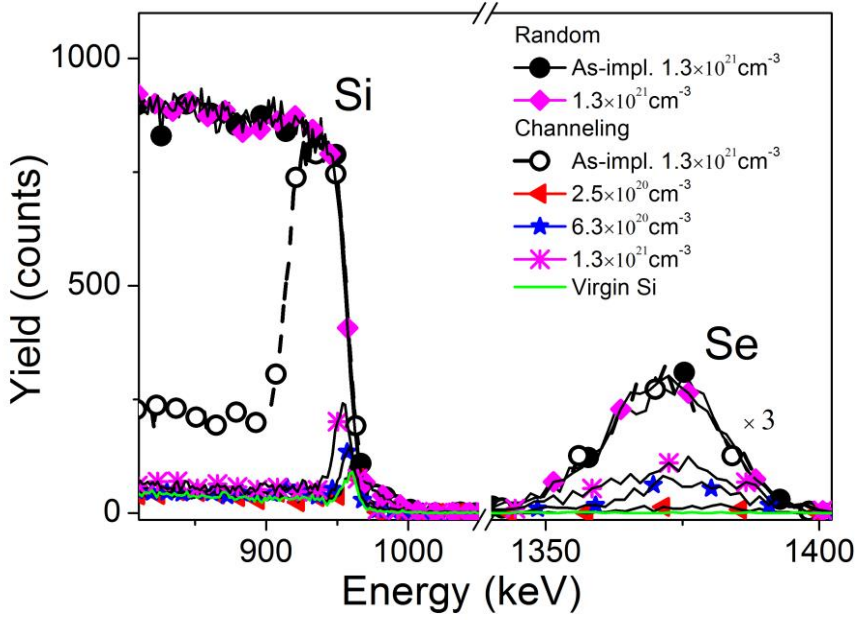

FIG. 2. RBS spectra of Se implanted Si samples with peak concentration of $2.5 \times 10^{20}$, $6.3 \times 10^{20}$, and $1.3 \times 10^{21}$ cm$^{-3}$ after FLA at 1.3 ms. The spectra for virgin Si are also included for comparison.

In Fig. 3, we present the temperature-dependent sheet resistance over a temperature range of 2.5–300 K for samples with different implantation fluences. A dramatic difference was observed in the sheet resistance for different samples. The sheet resistance of the as-implanted sample ($1.3 \times 10^{21}$ cm$^{-3}$) is added for comparison. The sheet resistance of the sample implanted with $2.5 \times 10^{20}$ cm$^{-3}$ Se sharply increases at low temperatures, viz. by up to 9 orders of magnitude compared to the sample with higher Se peak concentration ($1.3 \times 10^{21}$ cm$^{-3}$). The conductivity of this sample is thermally activated. The sample with the Se peak concentration of $6.3 \times 10^{20}$ cm$^{-3}$ shows finite conductivity as the temperature approaches zero and exhibits a metallic-like behavior. However, the sheet resistance of this sample slightly increases at low temperatures (see inset in Fig. 3). The sample with the Se peak concentration of $1.3 \times 10^{21}$ cm$^{-3}$ has the lowest sheet resistance and its value does not increase at low temperatures (see inset in Fig. 3). The insulator-to-metal transition in FLA-processed samples occurs at a Se peak concentration around $6.3 \times 10^{20}$ cm$^{-3}$. From DFT calculations performed by Ertekin *et al.* [14] a critical Se concentration of 1/249 (around $2.0 \times 10^{20}$ cm$^{-3}$) has been obtained. On the other hand, QMC calculations conducted by the same group predict a critical concentration of 1/128 (around $3.9 \times 10^{20}$ cm$^{-3}$). The relatively large discrepancy implies that the electron correlation effect should be considered. Additionally, Shao *et al.* [28] have employed a hybrid functional method for Se-hyperdoped Si and predicted a critical concentration of $5.09 \times 10^{20}$ cm$^{-3}$ at which the defect states merge into the conduction band. Our experimental results are in line with the critical values obtained by different theory approaches.

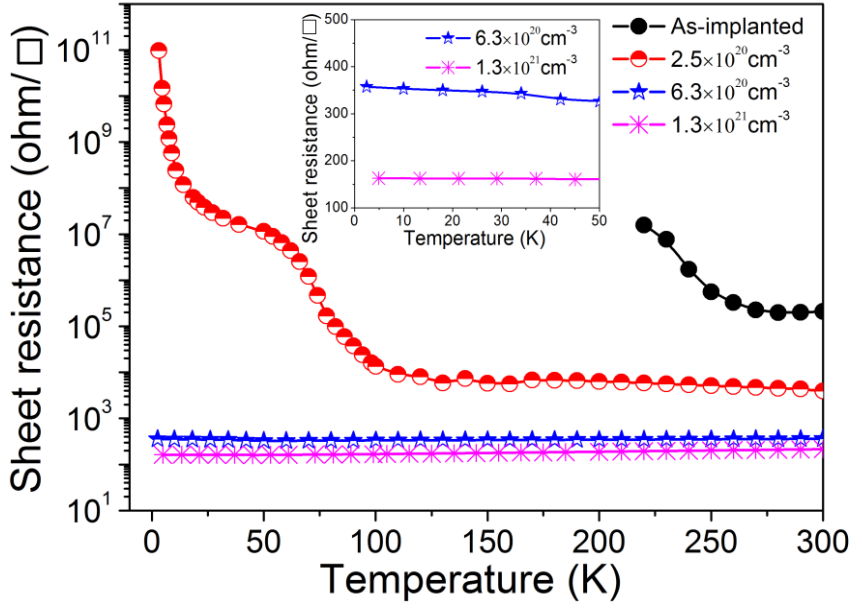

FIG. 3. Temperature-dependent sheet resistance of Se-hyperdoped Si after FLA. The result for the as-implanted sample ($1.3 \times 10^{21}$ cm$^{-3}$) is also shown for comparison. The inset shows the low temperature sheet resistance for samples with Se concentrations of $6.3 \times 10^{20}$ cm$^{-3}$ and $1.3 \times 10^{21}$ cm$^{-3}$.

According to Hall Effect measurements at room temperature, we calculate the carrier concentration by taking into account the effective thickness ($d_{eff}$) defined as $d_{eff} = \Phi/c_{pk}$ [8], which is 40 nm for all samples. The deduced carrier concentrations are $5.1 \times 10^{18}$, $7.3 \times 10^{19}$, and $9.0 \times 10^{19}$ cm$^{-3}$ for samples with Se peak concentration of $2.5 \times 10^{20}$, $6.3 \times 10^{20}$, and $1.3 \times 10^{21}$ cm$^{-3}$, respectively. So apparently the activation fraction (the ratio between the carrier concentration and the Se concentration) is only between 2% and 11%, a fact which is the subject of further investigation. The Mott criterion can be calculated from the equation $n_C^{1/3} a_B = 0.26 \pm 0.05$. The effective Bohr radius of $a_B$ is around 0.3 nm calculated by using the binding energy 200 meV for Se in Si [6, 7]. Thus, the critical concentration $n_C$ is around $6 \times 10^{20}$ cm$^{-3}$, which is larger than the measured carrier concentration for the metallic-like sample. The discrepancy can be due to the low-estimated carrier concentration from the experimental data, i.e., the peak carrier concentration can be higher. It also can be due to the shrinking of the binding energy of Se in Si with increasing the impurity concentration. According to the results mentioned above, we can conclude that the insulator-metal transition occurs in Se-hyperdoped Si fabricated by ion implantation and FLA. The conductivity of the sample with the Se peak concentration of $6.3 \times 10^{20}$ cm$^{-3}$ is around 700 /(ohm·cm) at low temperatures (2.5 K), which is comparable with the value of 600 – 800 /(ohm·cm) for the laser annealed sample with the Se peak concentration of $1.0\times 10^{21}$ cm$^{-3}$, reported in Ref. 14.

In order to gain further insight into the underlying physics behind the electrical properties of our samples, an experimental data modelling was performed by taking into account the Mott's theory about conduction mechanisms in insulating materials [8]. According to this theory, the conduction of Mott insulators proceeds by electron hopping between localized states at low-temperatures [31]. The insulating samples can therefore be modelled in terms of the conductivity as a function of temperature as follows:

$$\sigma(T) = \sigma_0 \exp[-(T_0/T)^s], \quad (1)$$

being $\sigma_0$ a prefactor, $T_0$ the activation temperature and s an exponent that defines the different hopping transport regimes. In detail, $s = 1$ corresponds to nearest-neighbor hopping, $s = ½$ implies a variable range hopping with a Coulomb gap in the density of states (DOS) and $s = ¼$ stands for a Mott's variable range hopping [32].

Our experimental data were fitted by using Eq. (1) with the exponent of $s = 1$, 1/2, and 1/4, respectively. Only $s = 1/2$ exponent was found to provide a good agreement at temperatures above 10 K as seen in Fig. 4. The activation temperature $T_0$ is also deduced to be 519 K (44 meV).

Alternatively, Zabrodskii [31] proposed a graphical differentiation procedure to readily identify the different hopping transport regimes. This lies on introducing a W (T) defined as

$$W(T) = \frac{d\ln(\sigma)}{d\ln(T)} \quad (2)$$

Hence, the equation (3) is obtained by inserting (1) into (2):

$$W(T) = s\left(\frac{T_0}{T}\right)^s \quad (3)$$

From this equation, the value of *s* can easily be inferred from the slope of the Log (W) - Log (T) representation. Based on this, a value of $0.53 \pm 0.01$ is obtained as shown in inset in Fig. 4. This suggests the presence of a Coulomb gap at the Fermi level giving rise to a variable range hopping mechanism [32]. The Coulomb gap vanishes as the Se-hyperdoped Si layers become more conducting.

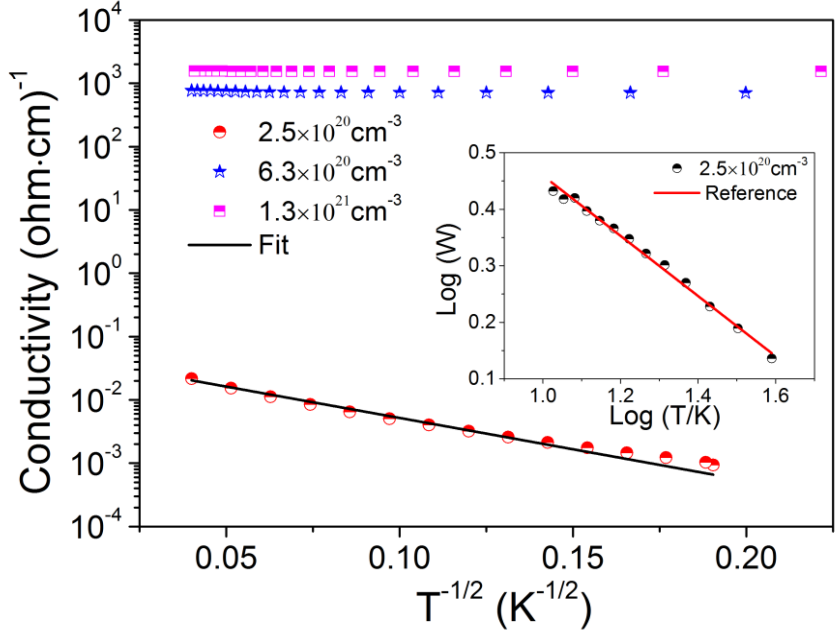

FIG. 4. Conductivity of the semi-insulating sample as a function of $T^{-1/2}$: the solid line is a fit of the experimental data to Eq. (1) with $s = 1/2$. The fitting range is 10 K< T< 50 K. The fit for the insulating sample yields $s = 0.53 \pm 0.01$. The inset shows the value of $s$ identified by the slope of $W(T) = \frac{d\ln(\sigma)}{d\ln(T)}$ versus T on a log scale. The red solid line shows $s = 0.50$ for reference.

## 4. Conclusions

We have investigated the structural and electrical properties of single-crystalline Si implanted with Se followed by FLA. From Raman and RBS/C results, it has been demonstrated that high crystalline quality of the implanted samples has been achieved after FLA. Increasing the Se doping concentration from $2.5 \times 10^{20}$ to $6.3 \times 10^{20}$ cm$^{-3}$ (less than 3 times) resulted in a conductivity enhancement of several orders of magnitude, leading to metallic-like behavior and demonstrating an insulator-to-metal transition among these values. This is in quantitative agreement with the values predicted from DFT and QMC calculations. Based on the temperature-dependent electrical measurements, the semi-insulating samples have revealed variable-range-hopping conduction with a Coulomb gap at the Fermi level. It is worth to note that the whole processing (ion implantation + flash lamp annealing) is compatible with the Si-based chip technology and can be easily integrated with the existing production line.

## Acknowledgement

Support by the Ion Beam Center (IBC) at HZDR is gratefully acknowledged. This work is supported by the Project Helmholtz-Gemeinschaft Deutscher Forschungszentren (HGF-VH-NG-713).

F. L. is supported by the China Scholarship Council (File No.201307040037). Y. B. would like to thank the Alexander-von-Humboldt foundation for providing a postdoctoral fellowship.